\documentclass[twocolumn,showpacs,showkeys]{revtex4}
\usepackage{float}
\usepackage{graphics}
\usepackage{amsmath}
\usepackage{SIunits}
\usepackage{verbatim}

\floatstyle{boxed}
\newfloat{boxflt}{p}{lob}
\floatname{boxflt}{BOX}

\newcommand{\ben}{\begin{equation}}
\newcommand{\een}{\end{equation}}
\newcommand{\bean}{\begin{eqnarray}}
\newcommand{\eean}{\end{eqnarray}}

\newcommand{\bea}{\begin{eqnarray*}}
\newcommand{\eea}{\end{eqnarray*}}

\newcommand{\latin}[1]{\textit{#1}}
\newcommand{\dif}{\mathrm{d}}

\addunit{\byte}{B}
\addunit{\mebi}{Mi}

\begin{document}


\title{Probing the Effects of the Well-mixed Assumption on Viral Infection Dynamics}
\author{Catherine Beauchemin\footnote{Electronic address: \texttt{cbeau@phys.ualberta.ca}}}
\affiliation{Department of Physics, University of Alberta, Edmonton, AB, T6G 2J1, Canada}
\date{\today}


\begin{abstract}
Viral kinetics have been extensively studied in the past through the use of spatially well-mixed ordinary differential equations describing the time evolution of the diseased state. However, emerging spatial structures such as localized populations of dead cells might adversely affect the spread of infection, similar to the manner in which a counter-fire can stop a forest fire from spreading. In a previous publication \cite{cbeau05}, a simple 2-D cellular automaton model was introduced and shown to be accurate enough to model an uncomplicated infection with influenza A. Here, this model is used to investigate the effects of relaxing the well-mixed assumption. Particularly, the effects of the initial distribution of infected cells, the regeneration rule for dead epithelial cells, and the proliferation rule for immune cells are explored and shown to have an important impact on the development and outcome of the viral infection in our model.
\end{abstract}



\maketitle

\section{Introduction}

Mathematical modelling of viral infection dynamics has become a very popular approach to understanding and characterizing the dynamics of viral infections. The basic viral infection model, which was introduced by Perelson \cite{perelson02,perelson96hiv}, namely
\bea
\frac{\dif T}{\dif t} &=& \lambda - d T - k T V \\
\frac{\dif I}{\dif t} &=& k T V  - \delta I \\
\frac{\dif V}{\dif t} &=& p I - c V
\eea
describes the temporal evolution of the population of susceptible or target cells, $T$, which become infected, $I$, as a result of their interactions with virus particles, $V$. This model is widely used with minor or major modifications to study the dynamics of various viral infections. Typically, these mathematical modelling efforts seek to determine crucial parameters of the dynamics of a specific viral infection which would be impractical or arduous to extract experimentally.

But those simple ODE models make the very important assumption that the various populations of cells and virions are uniformly distributed over the space where the infection takes place for all times; an assumption that is rarely realistic, and which may or may not affect in a significant way the resulting dynamics. For this reason, there is growing interest in probing the effect of spatial distribution on systems in ecology \cite{durrett94,durrett-levin94,young01}, epidemiology \cite{lloyd96,hagenaars04} and immunology \cite{funk05,louzoun01,strain02}.

Here, I explore the effects of spatial structures on the dynamics of a viral infection, whose target cells are fixed in space, using a 2-D cellular automaton introduced in previous work \cite{cbeau05}. I will explore which kind of effects spatial structures can have on the evolution and outcome of a spatially localized viral infection. I will also show how these spatial structures emerge and by which process they affect the dynamics of the infection.

In the next section the reader will briefly be reminded about the rules and parameters of the cellular automaton model. Then, in Section \ref{sec_patch}, the effect of the distribution of initially infected cells on the progression of the infection is investigated. Section \ref{sec_regen} compares a local regeneration rule for epithelial cells to a global rule, i.e.\ the rule for the replacement of dead epithelial cells with healthy cells. In Section \ref{sec_imm}, the effects of the addition of immune cells at random locations versus addition at the site of recruitment are explored. Finally, in Section \ref{sec_flu}, the significance of the spatial effects in the particular case of an uncomplicated influenza A viral infection is discussed.

\section{The cellular automaton}

The cellular automaton (CA) model that will be used in this work was introduced in \cite{cbeau05}, where the values chosen for each parameter are also justified. The relationship between the notation used here and that of \cite{cbeau05} is listed in the Appendix. The model considers 2 species of cells: epithelial cells which are the target of the viral infection, and immune cells which fight the infection. The CA is run on a 2-D square lattice where each site represents one epithelial cell, and immune cells are mobile, moving from one lattice site (epithelial cell) to another. The simulation grid is updated synchronously and has toroidal boundary conditions for both cell types. The virus particles are not explicitly considered, rather the infection is modelled as spreading directly from one epithelial cell to another.

The evolution rules of the CA model for the epithelial and immune cell species are enumerated in Boxes \ref{t_rules_epithelial} and \ref{t_rules_immune}, respectively. At initialization time, each epithelial cell is assigned a random age between 0 and $\delta_H$. All but a fraction $\rho_C = 0.01$ epithelial cells are initialized as healthy, the rest are set as containing virions. Additionally, $\rho_M = 1.5\times10^{-4}$ unactivated immune cells per epithelial cell are placed at random locations on the grid.

\begin{boxflt}
\resizebox{0.9\linewidth}{!}{\includegraphics{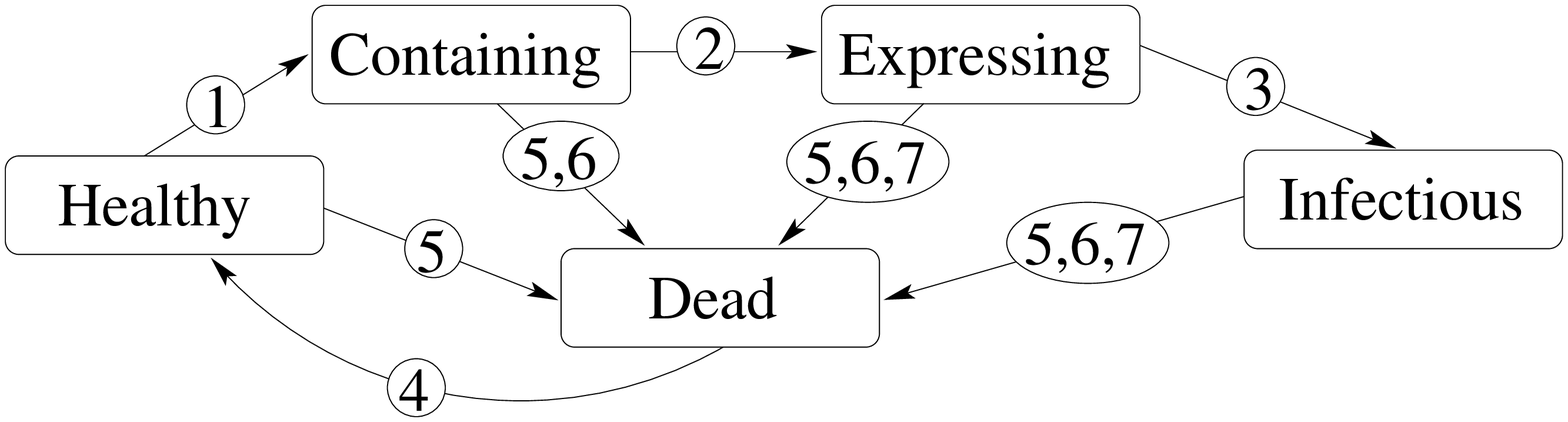}} \\
\begin{enumerate}
\item Healthy epithelial cells get infected with probability $\beta = \unit{2}{\hour^{-1}}$ for each infectious Moore neighbour (8 nearest neighbours).
\item A cell containing virions and which has been infected for $\tau_E = \unit{4}{\hour}$ begins expressing the viral peptide on their epitope.
\item An expressing cell that has been expressing its viral peptide for $\tau_I = \unit{2}{\hour}$ becomes infectious.
\item A dead cell becomes a healthy cell (i.e.\ is replaced by a healthy cell) if any of its 8 Moore neighbours divides. Only healthy cells divide and they divide every $b = \unit{12}{\hour}$.
\item All cells will die of old age after living for exactly $\delta_H = \unit{380}{\hour}$, unless they die earlier because of viral toxicity or immune recognition (see below).
\item Because of viral toxicity, infected cells (i.e.\ containing+expressing+infectious) will die after having been infected for $\delta_I = \unit{24}{\hour}$, unless they die earlier from recognition (see below) or from old age (see above).
\item Finally, expressing and infectious cells die when ``recognized'' by an activated immune cell.
\end{enumerate}
\caption{Evolution rules for the epithelial cells in the cellular automaton model.}
\label{t_rules_epithelial}
\end{boxflt}

\begin{boxflt}
\resizebox{0.9\linewidth}{!}{\includegraphics{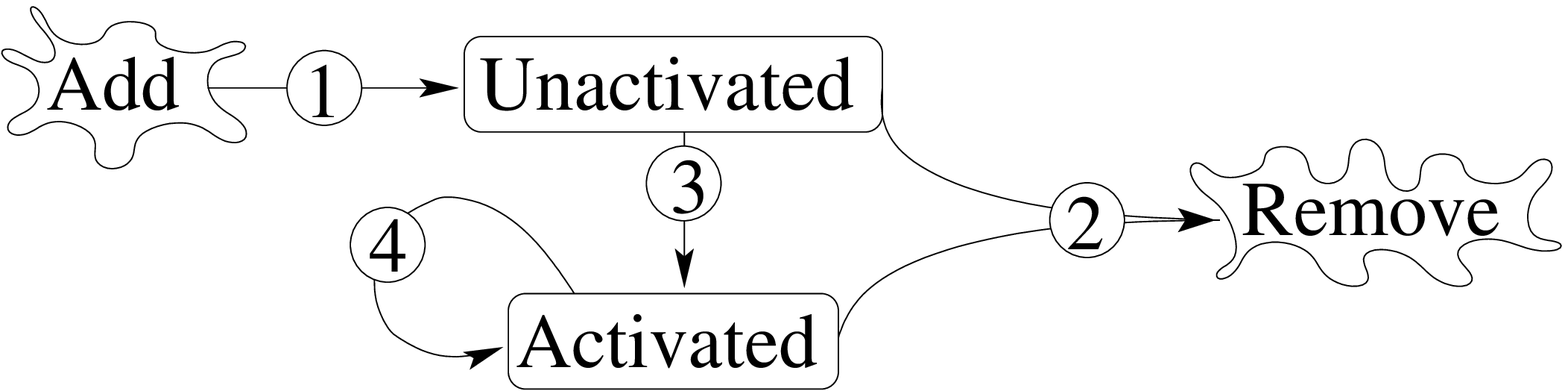}} \\
\begin{enumerate}
\item Unactivated immune cells are added at random lattice sites as needed to maintain a minimum density of $\rho_M = 1.5\times10^{-4}$ unactivated immune cells.
\item All immune cells die of old age after living for exactly $\delta_M = \unit{168}{\hour}$.
\item An unactivated immune cell becomes activated when it first occupies an expressing or infectious lattice site.
\item If an activated cell is occupying an expressing or infectious lattice site, it kills the epithelial cell and with a probability $r_M = 0.25$ a new activated immune cells is added at a random location on the grid.
\end{enumerate}
Additionally, immune cells move randomly on the CA lattice at a speed of one lattice site per time step, and there are $\nu = \unit{6}{time\ steps/\hour}$.
\caption{Evolution rules for the immune cells in the cellular automaton model.}
\label{t_rules_immune}
\end{boxflt}

\section{Distribution of initially infected cells}
\label{sec_patch}

In our CA model, the parameter $\rho_C$ is the fraction of epithelial cells initially set in the infected state, and its default value is 1\%. In \cite{cbeau05}, the cells to be initially set to the infected state were picked at random and this resulted in single infected cells as well as groupings or patches of neighbouring infected cells of various sizes. One way to investigate the effect of spatial heterogeneities on the dynamics of the infection is to change the spatial configuration of the epithelial cells that are initially set in the infected state. To do this, our model was modified to distribute the initially infected cells into groups or patches of fixed size so that the effect of the size of the patches of infected cells on the dynamics of the infection can be investigated.

A new parameter, $s$, is added to our CA model, being the number of cells that make up a patch of initially infected cells. Since the number of epithelial cells to be initially infected is not necessarily divisible by $s$, the quotient of that division gives the number of patches to be added to the simulation grid at start up, and the remainder of the division is used to set the probability that an extra patch of size $s$ be added. This means that a fixed initial patch size is enforced at the expense of a fixed fraction of initially infected cells. Each patch of infected cells is individually constructed and is added at a random location on the grid, insuring that no two patches are in contact with each other. Our model defines the neighbourhood of a site as consisting of the site itself and its eight closest sites \cite{cbeau05} (Moore neighbourhood). A patch of $s$ infected cells is constructed by starting with a seed site and growing it by sequentially picking one site at random from the set of sites that neighbour previously-selected sites. Note that this method of forming patches results in patches with densities that decrease with increasing distance from the centre. This characteristic is consistent with a splatter or spray of virions and thus this method was preferred over other patch growing methods such as diffusion-limited aggregation, and random walk additions around a seed.

\begin{figure*}
\begin{center}
\resizebox{0.45\linewidth}{!}{\includegraphics{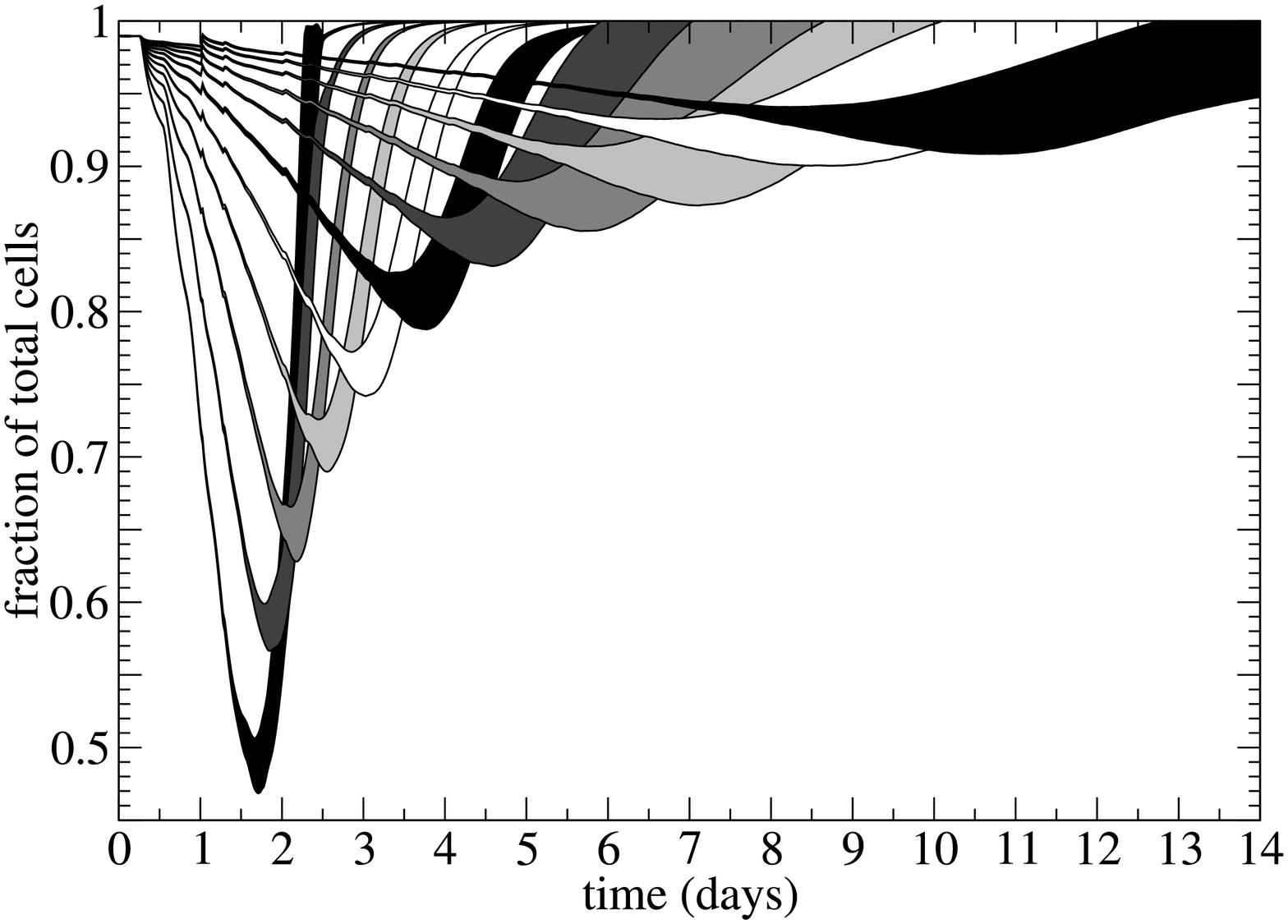}}
\hspace{0.01\linewidth}
\resizebox{0.45\linewidth}{!}{\includegraphics{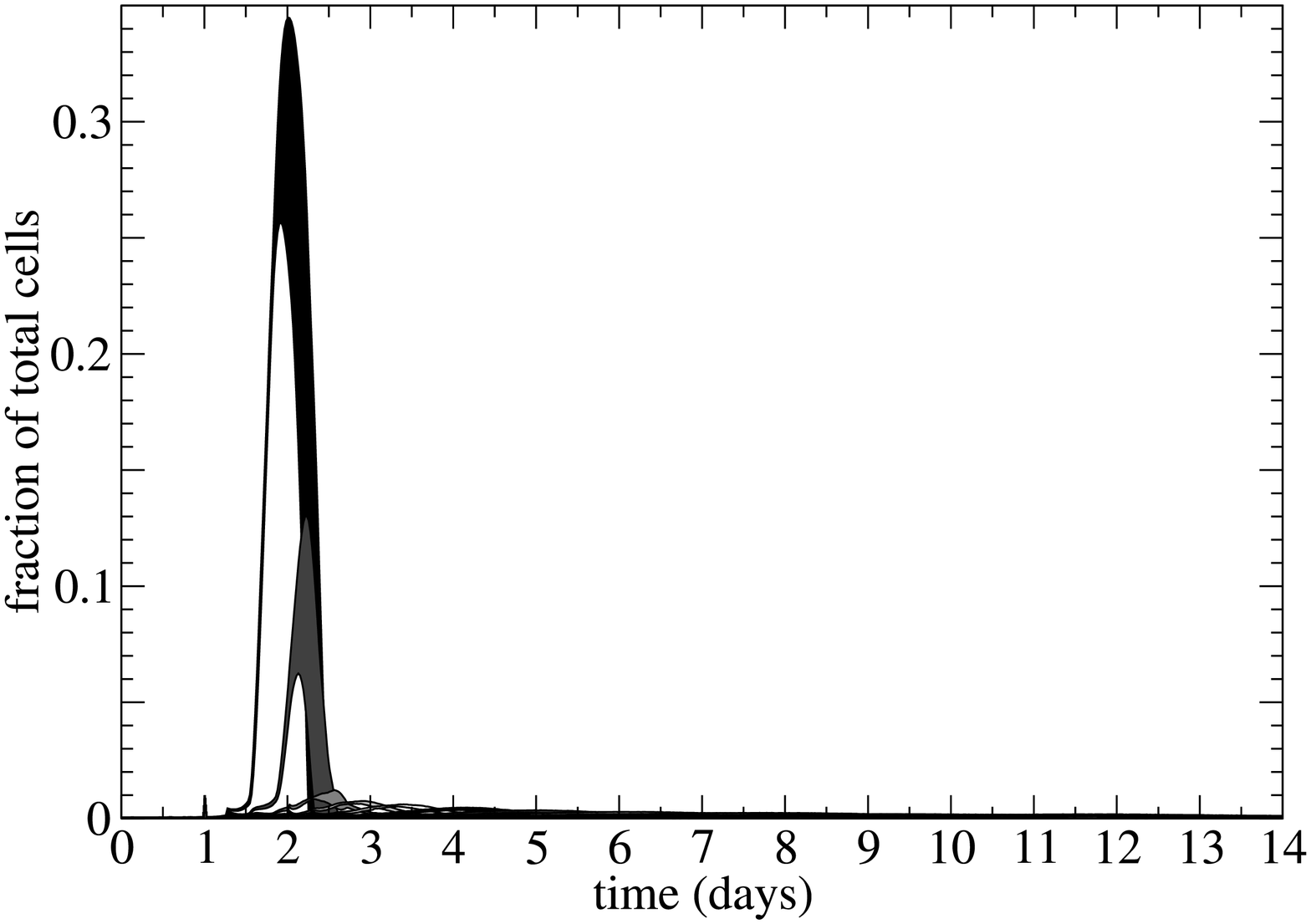}} \\
\vspace{0.01\linewidth}
\resizebox{0.45\linewidth}{!}{\includegraphics{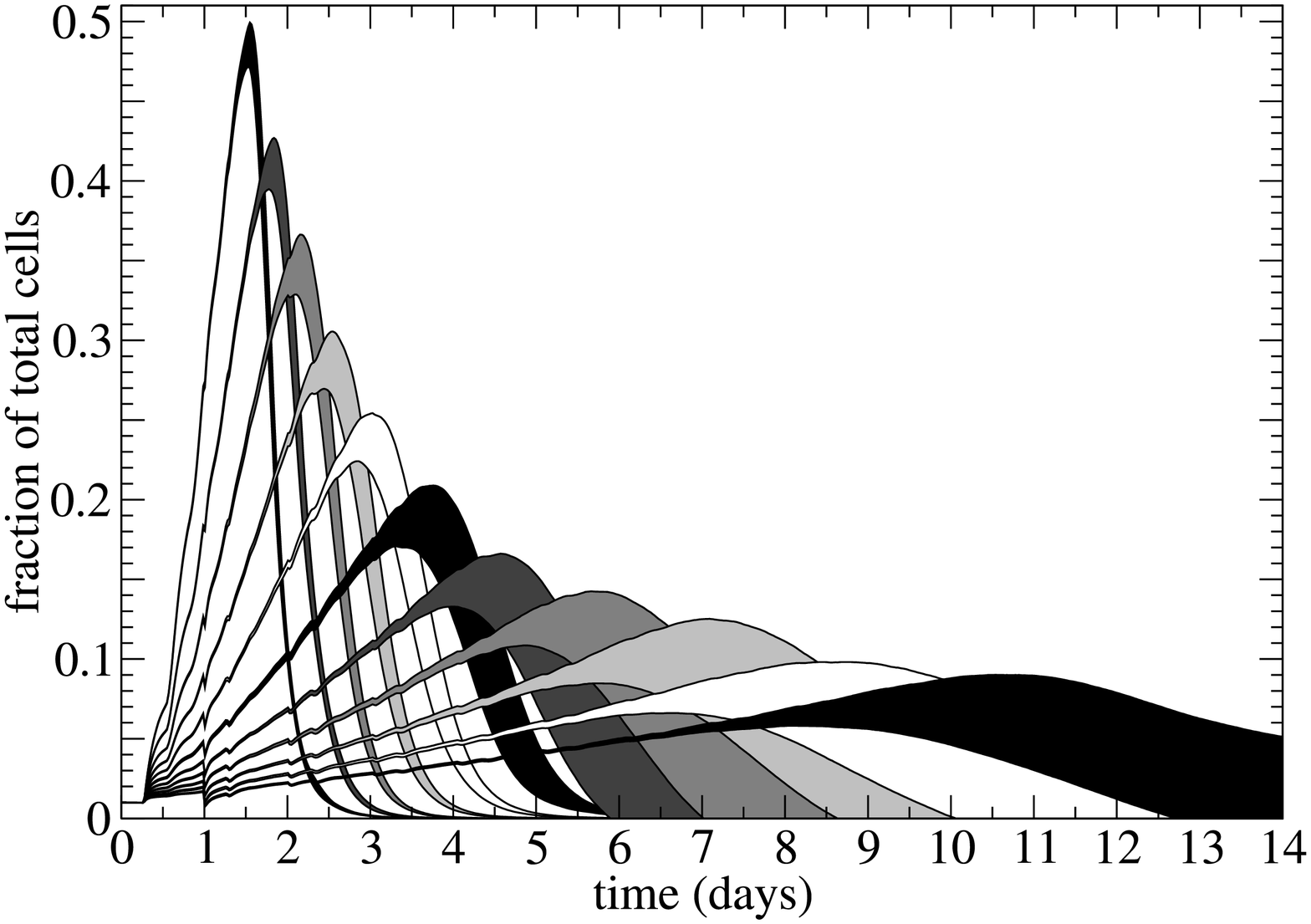}}
\hspace{0.01\linewidth}
\resizebox{0.45\linewidth}{!}{\includegraphics{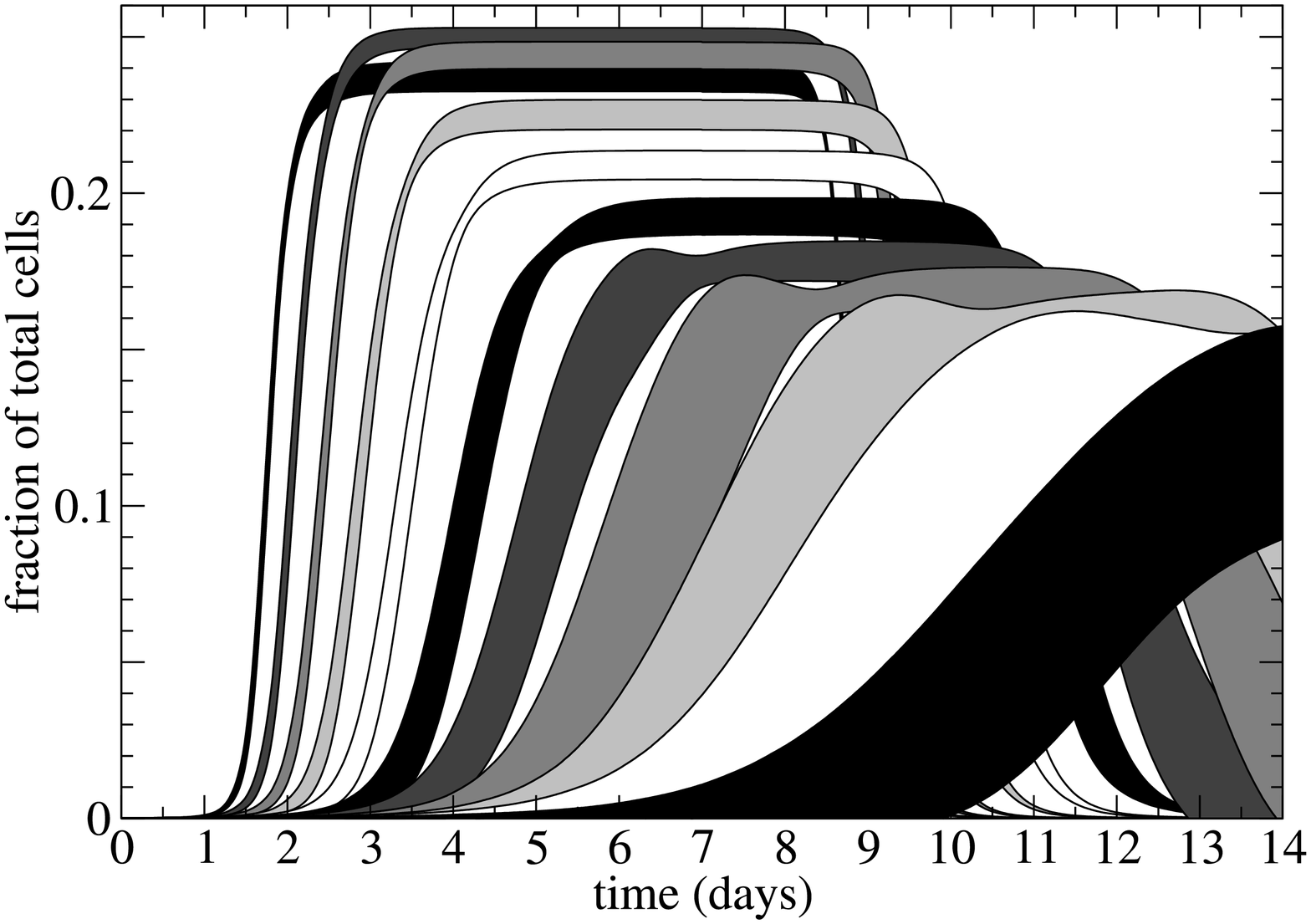}}
\end{center}
\caption{The effect of varying the initial patch size, $s$, on the viral infection's dynamics. The graphs show the time evolution of the populations of healthy (top left), dead (top right), infected (bottom left), and immune cells (bottom right) for $s$ values of $1$, $2$, $4$, $8$, $16$, $35$, $77$, $154$, $308$, $616$, and $1232$ cells. The greyed areas mark one standard deviation after 50 runs for each initial patch size, with periodically decreasing darkness corresponding to increasing initial patch sizes. In all cases, the black band that peaks first is $s = 1$. The graphs show that the dynamics of the viral infection is sensitive to the spatial organization of the initially infected epithelial cells.}
\label{g_patch}
\end{figure*}

The results for patches ranging in size from 1 to 1232 infected cells are presented in Figure \ref{g_patch}.
One can see that increasing initial patch sizes result in fewer infected cells and less epithelial damage. This is not surprising since only the cells that make up the perimeter of the patch, i.e.\ those that have healthy neighbours, can infect other cells. As patches grow, their perimeter to area ratio, namely the fraction of infectious cells that have healthy neighbours, will decrease and so will the effective infection rate.

\begin{figure}
\begin{center}
\begin{tabular}[b]{|c|c|c|}
\hline
Time & \# inf.\ & \# new inf.\ \\
\hline
0 & 1 & 8 \\
1 & 9 & 16 \\
2 & 25 & 24 \\
3 & 49 & 32 \\
\vdots & \vdots & \vdots \\
$n$ & $(2n+1)^2$ & $8(n+1)$ \\
\hline
\end{tabular}
\resizebox{0.45\linewidth}{!}{\includegraphics{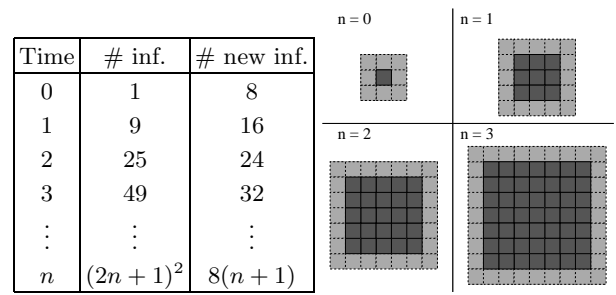}}
\end{center}
\caption{Evolution of a simplified system where each infected cell infects all of its uninfected neighbours at each time step, starting from a single infected cell. The table shows the number of infected cells and the number of cells that will become infected in the next time step. The figure illustrates the evolution of the system over the first 4 time steps with infected cells represented in dark grey and the cells which will be infected in the next time step represented in light grey.}
\label{g_inf_example}
\end{figure}

\begin{figure}
\begin{center}
\resizebox{0.9\linewidth}{!}{\includegraphics{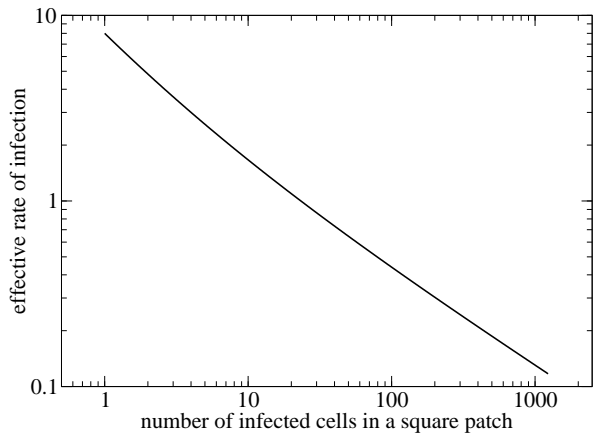}}
\end{center}
\caption{The effective rate of infection (newly infected cells per infected cell) as a function of number of infected cells in a square patch for the simplified system presented in Figure \ref{g_inf_example}. The effective infection rate is given by $4/\sqrt{I} + 4/I$ where $I$ is the number of infected cells that make up the square patch.}
\label{g_eff_inf}
\end{figure}

Let us illustrate this by an example. Consider a system where infected cells infect all of their uninfected Moore neighbours (8 nearest neighbours) in each time step (an infection rate of $100\%$). The evolution of the system from an initial single seed is illustrated in Figure \ref{g_inf_example}.
From the relation derived in the table of Figure \ref{g_inf_example}, one can compute the effective infection rate, i.e.\ the number of newly infected cells per infected cell at time step $n$, to be $8(n+1)/(2n+1)^2 = 4/\sqrt{I} + 4/I$, where $I = (2n+1)^2$ is the number of infected cells in a square patch at time step $n$. A graph of the effective infection rate as a function of the number of infected cells in a square patch is presented in Figure \ref{g_eff_inf}.
For this toy model, the effective infection rate is proportional to $1/\sqrt{I}$ for $I \gg 1$.

\begin{figure}
\begin{center}
\resizebox{0.9\linewidth}{!}{\includegraphics{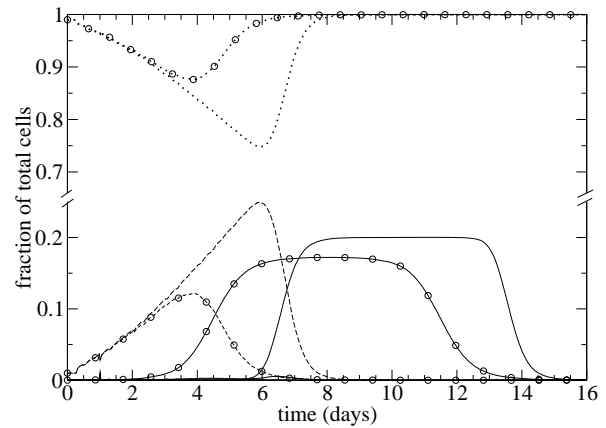}}
\end{center}
\caption{Proportion of healthy cells (dotted), infected cells (dashed), and immune cells per epithelial cell (full) for two simulations using an initial patch size of $s = 77$. The simulations, whose only difference is the seed for the random number generator, illustrate the differences that can arise for large values of the initial patch size. In this case, early immune detection (lines with circles) of the infection has allowed minimal damage and early recovery, while late detection (lines without symbols) has resulted in a longer infection with a larger number of infected and dead cells.}
\label{g_variability}
\end{figure}

Another interesting feature that can be seen in Figure \ref{g_patch} is the increasing standard deviation for increasing initial patch sizes. This is easily explained with the fact that the larger the parameter $s$, the fewer the sites of infection. In other words, as the initial patch size increases, the 50 simulations are averaging over fewer infection sites. Figure \ref{g_variability} presents two example simulations to illustrate the differences that can arise between simulations produced using the same parameter values, when the initial patch size is large. In the case of the example simulations presented in Figure \ref{g_variability}, early detection made the difference between a small and short infection, and a longer infection resulting in a greater number of infected and dead cells. The larger the initial patch size, the fewer the number of infected patches and thus, the more pronounced this effect will be. This variability for larger values of $s$ can be reduced by averaging simulations with the same number of infection sites (same number of patches) rather than the same absolute number of infected cells (same area).

\subsection{Not just a rescaling problem}

\begin{figure}
\begin{center}
\resizebox{0.9\linewidth}{!}{\includegraphics{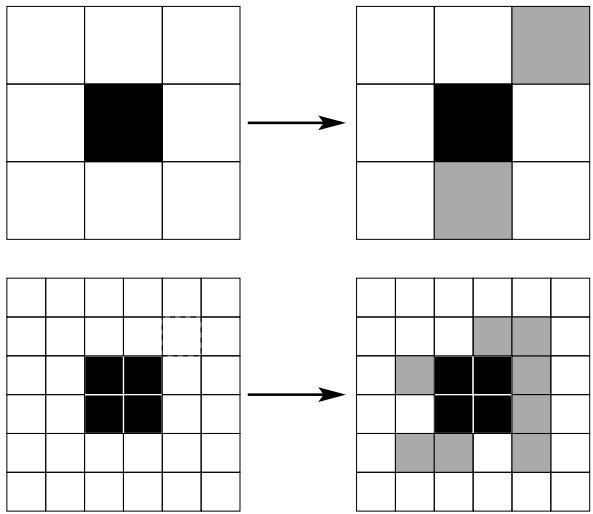}}
\end{center}
\caption{The comparison of the infection growth pattern for a simulation where each epithelial cell is represented by: (top) a single grid site; or (bottom) 4 grid sites ($s = 4$). For the infection growth rate to be comparable for the 2 simulations, the fraction of the grid which gets infected needs to be kept constant such that an infection rate $\beta$ for an initial patch size of $1$ becomes $\beta*s$ for an initial patch size of $s$. Despite this correction, the infection growth pattern is not equivalent because, for example, the radius of the infection increases faster in the former.}
\label{g_growth}
\end{figure}

It may be tempting to interpret the effect of the initial patch size on the development and outcome of the infection as a rescaling of the system. In effect, one could imagine that each lump of infected cells represents a single infected cell such that the surface area of one epithelial cell corresponds to $s$ sites of the simulation grid. A grid of area $A$ with an initial patch size of $s$ would be equivalent to a grid of area $A/s$ with an initial patch size of 1. This turns out to be an incorrect interpretation, as seen in Figure \ref{g_growth}. This figure illustrates that one consequence of increasing the number of simulation sites per epithelial cell is an increase in the number of configurations the simulation can be in. For example, this causes the radius of infection sites to grow more slowly, even when the cell-to-cell infection rate is increased so that the rate of increase of infected tissue area is kept constant.

\subsection{Occurrence of chronic infection}
It is not clear from Figure \ref{g_patch}, but for initial patch sizes larger than 35, a number of simulations result in chronic infection with the fraction of infected cells stabilizing at $2\%$ in all such cases. The occurrence of chronic infection increases for increasing initial patch sizes. This is illustrated in the top left panel of Figure \ref{g_chronic}. What causes chronic infections in the case of larger initial patch sizes is the lower effective infection rate, which slows the infection dynamics. If the infection growth is slowed down, the infection takes place over a longer period of time and the immune cells start dying off before the infection is fully cleared. Thus, in the CA model, chronic infection arises when the immune cells' lifespan is shorter than the time scale of the infection. Chronic infections can be prevented by choosing a larger value for $\delta_M$, the lifespan of immune cells, for larger values of $s$, the initial patch size. For $s = 1232$, there are still occurrences of chronic infection with $\delta_M = \unit{300}{\hour}$, but the infections are always cleared for $\delta_M = \unit{400}{\hour}$ (not shown).

\section{Global vs local epithelial regeneration}
\label{sec_regen}

In the model presented in \cite{cbeau05}, the regeneration of dead epithelial cells was implemented as a global process rather than a local process, namely, a dead cell is replaced by a healthy cell with probability $b^{-1} \times \text{\# healthy} / \text{\# dead}$. This epithelial cell regeneration rule was originally chosen to mimic the replacement of dead cells by basal cells or by cells from inferior layers in the context of an influenza A infection. If one, instead, considers an infection taking place in a tissue composed of a monolayer of cells, a local regeneration rule based on the division of immediate neighbours is more appropriate. In this section, the impact of using the local epithelial cell regeneration rule on the dynamics of the infection is investigated. The local regeneration rule is such that a dead epithelial cell is replaced by a healthy one only if one of its healthy neighbours divides.

The original global regeneration rule is equivalent to assuming that dead and healthy epithelial cells are homogeneously distributed throughout the simulation grid, which is the way in which epithelial regeneration is implemented in simple ODE models. Comparing the two regeneration rules allows us more insight into the effect of the spatial distribution of cells on localized infection dynamics. The results of simulations comparing the global to the local epithelial cell regeneration rules are shown in the left column of Figure \ref{g_simres}. The top left panel shows the original model with the global epithelial cell regeneration rule, as presented in \cite{cbeau05}, and the bottom left panel shows the same model using the local epithelial cell regeneration rule. A typical spatial distribution of cells at day 4 post-infection for both rules is illustrated in the left column of Figure \ref{g_sshot} as screenshots of the simulation grid, with the panels in the same order as in Figure \ref{g_simres}. Additionally, the numbers of infected and dead cells at their respective peaks relative to their values in the original CA model introduced in \cite{cbeau05} are presented in Table \ref{t_compare} in the two rows labelled ``newly recruited immune cells placed at random locations;''  the other rows will be discussed in the next section. One can see that the local epithelial cell regeneration rule results in fewer infected cells and, consequently, in the recruitment of fewer immune cells but in more extensive and longer lasting damage to the epithelium compared to the global regeneration rule.

In the CA model, the infection of epithelial cells spreads locally as infected cells infect their healthy neighbours forming growing patches of infected cells. As the infection progresses, infected cells at the core of these patches die as a result of virus toxicity or immune attacks, and leave behind patches of dead cells surrounded by a perimeter of infected cells. Patches of dead cells can no longer harbour infection and thus serve to limit the growth of the infection. With the global epithelial cell regeneration rule, new healthy cells are allowed to emerge in the middle of the pools of dead cells. This allows the infection to rapidly repopulate the patches of dead cells, thus sustaining a high level of infection with minimal epithelial damage.

With the local epithelial cell regeneration rule, the patches of dead epithelial cells can only be repopulated by healthy cells once the immune cells have begun destroying the rings of infected cells that encircle each patch of dead cells, which otherwise act as a barrier isolating healthy cells from the areas that require regeneration. Thus, the greater accumulation of damage that results from the use of the local regeneration rule is a consequence of the spatial constraints imposed on the regeneration process. This finding is in agreement with that of Strain et al.\ \cite{strain02}, who reported that for their spatial model of HIV, the infection could only be sustained as a propagating wave when the local rate of cell death was greater than the local regeneration rate, as is the case with our model when using the local regeneration rule for epithelial cells.

\subsection{Occurrence of chronic infection}

Examination of the results of the local epithelial cell regeneration rules for various initial patch sizes reveals the persistence of infected cells, namely a chronic infection stabilizing at approximately $1\%$ of cells infected, for all but an initial patch size of 1. This is illustrated in Figure \ref{g_chronic}, where the fraction of simulations ending in chronic infection as a function of the initial patch size for the local epithelial cell regeneration rule is presented in the bottom left panel. The smaller number of infected epithelial cells resulting from the use of the local regeneration rule results in the recruitment of fewer immune cells making it harder to fight the viral infection. Additionally, the organization of the infected cells into circular waves makes it harder for the immune cells to target the infected cells' structures. When infected cells are arranged into patches, an immune cell performing a random walk has better chances of landing on multiple infected sites. When infected epithelial cells organize into rings, as is the case with the local regeneration rule, immune cells performing a random walk will often move off the ring structure and ``lose sight'' of the infection. Consequently, the smaller number of infected cells and their organization into circular waves, facilitates the escape of the infection from immune attacks resulting in a higher incidence of chronic infections than for a global epithelial cell regeneration rule.

\section{Immune cells' proliferation rule}
\label{sec_imm}

\begin{figure*}
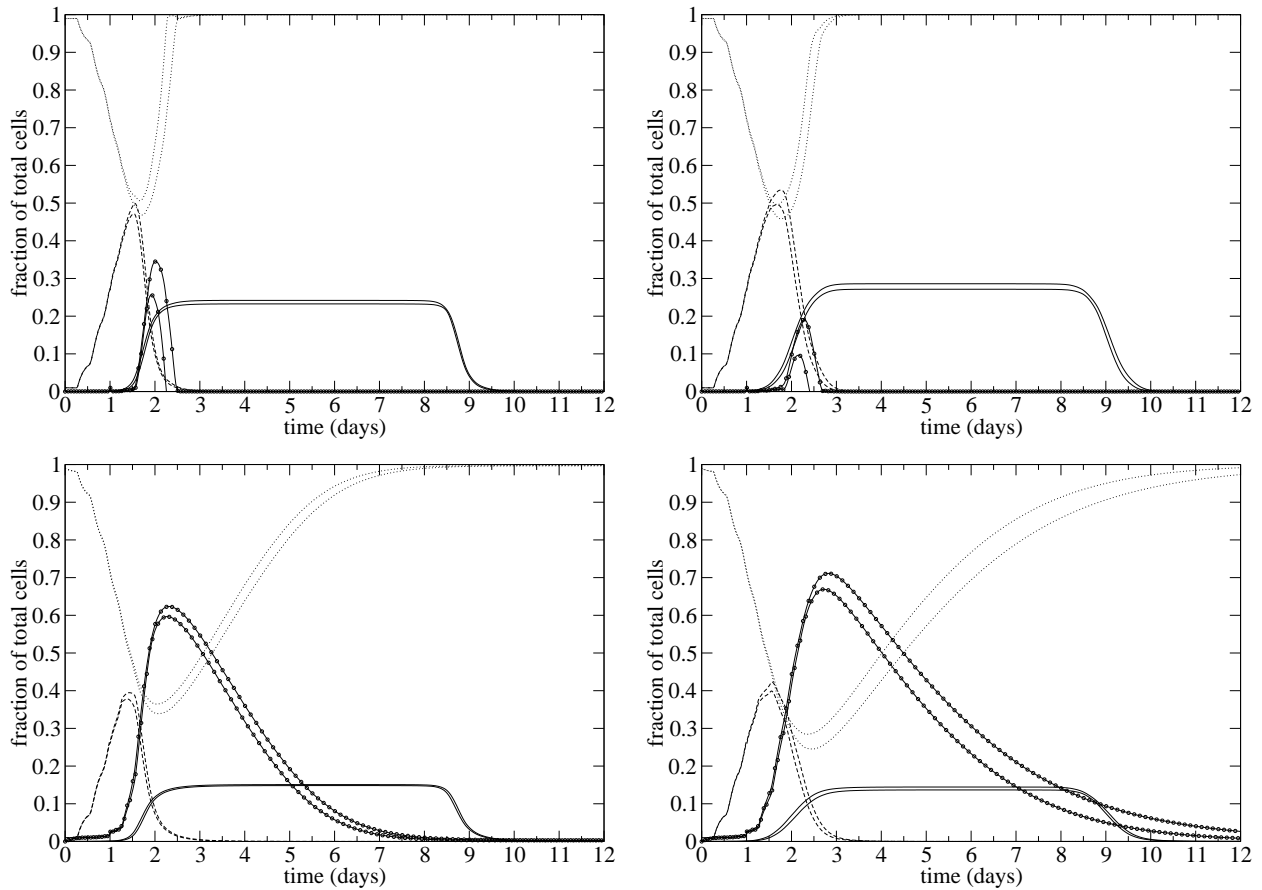

\begin{center}
\resizebox{0.45\linewidth}{!}{\includegraphics{original}}
\hspace{0.01\linewidth}
\resizebox{0.45\linewidth}{!}{\includegraphics{divide_on_mom}} \\
\vspace{0.01\linewidth}
\resizebox{0.45\linewidth}{!}{\includegraphics{local_regen}}
\hspace{0.01\linewidth}
\resizebox{0.45\linewidth}{!}{\includegraphics{local_on_mom}}
\end{center}
\caption{The effect of a global (top row) or local (bottom row) epithelial cell regeneration rule with the addition of immune cells at random sites (left column) or at the site of recruitment (right column) on the behaviour of the CA model. Simulation results averaged over 50 simulation runs for an initial patch size of 1. The paired lines mark one standard deviation and represent the fraction of epithelial cells that are healthy (dotted), infected (dashed), dead (full with circles), as well as the proportion of immune cells per epithelial cells (full). The top left panel corresponds to the original model presented in \cite{cbeau05}.}
\label{g_simres}
\end{figure*}

\begin{figure*}
\begin{center}
\fbox{\resizebox{0.48\linewidth}{!}{\includegraphics{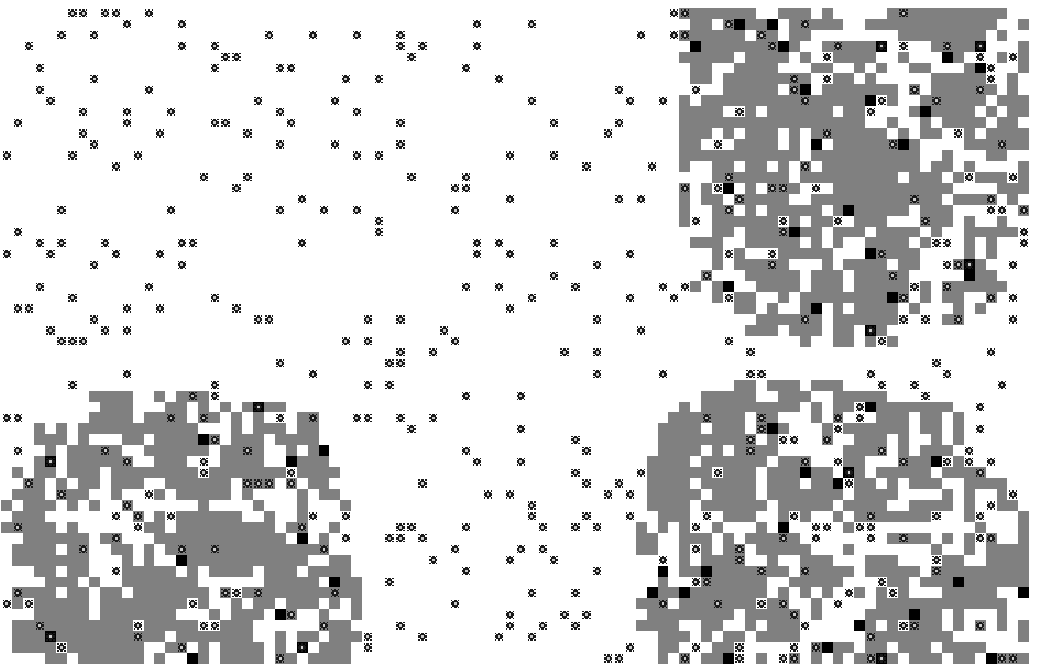}}}
\fbox{\resizebox{0.48\linewidth}{!}{\includegraphics{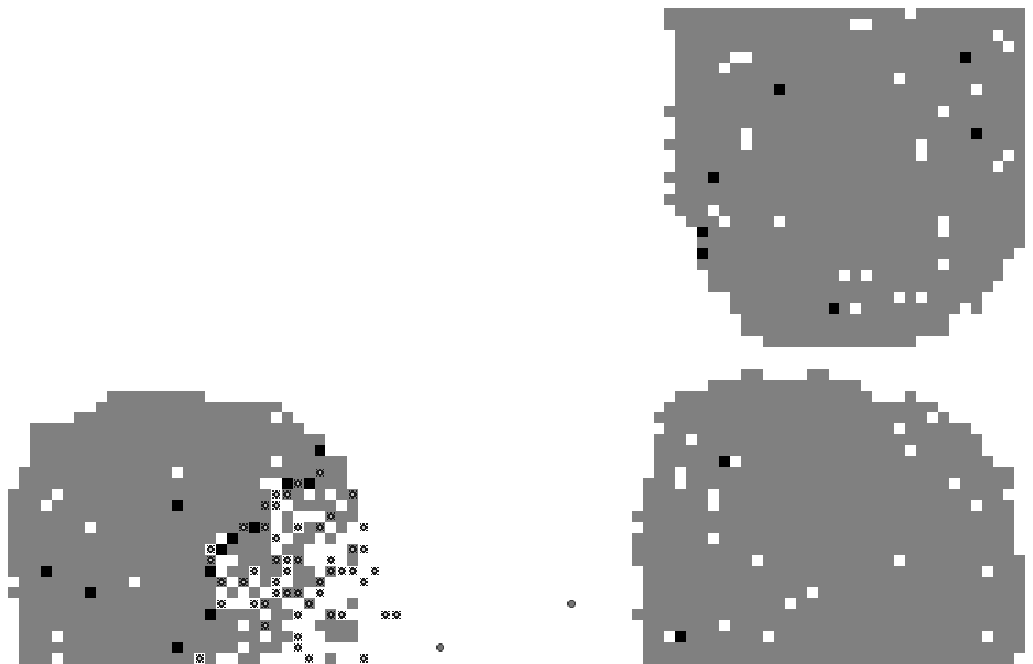}}}
\fbox{\resizebox{0.48\linewidth}{!}{\includegraphics{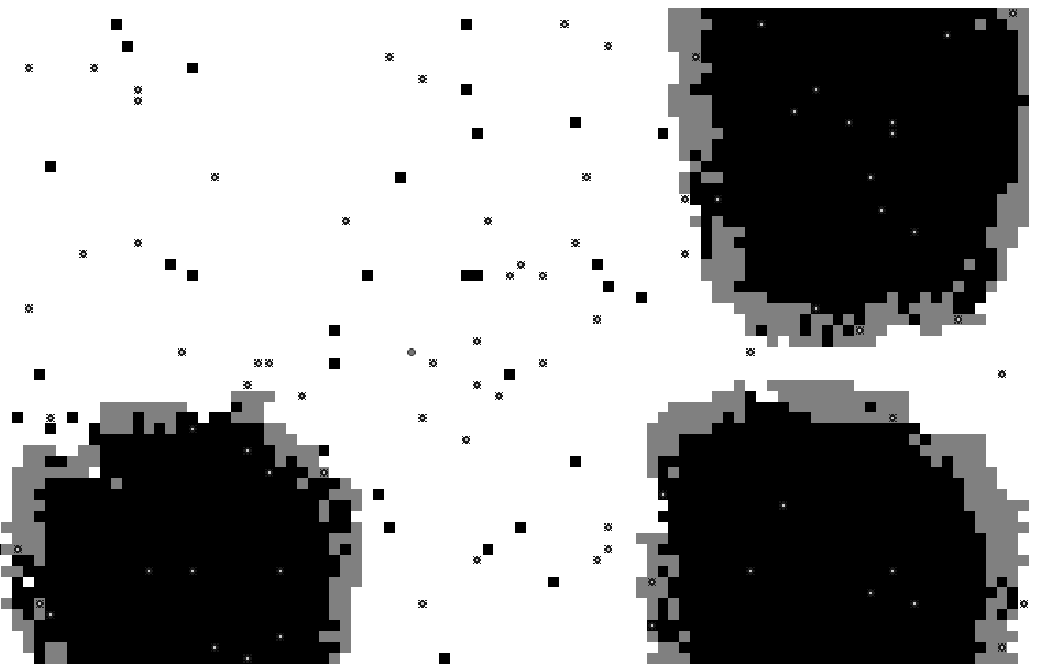}}}
\fbox{\resizebox{0.48\linewidth}{!}{\includegraphics{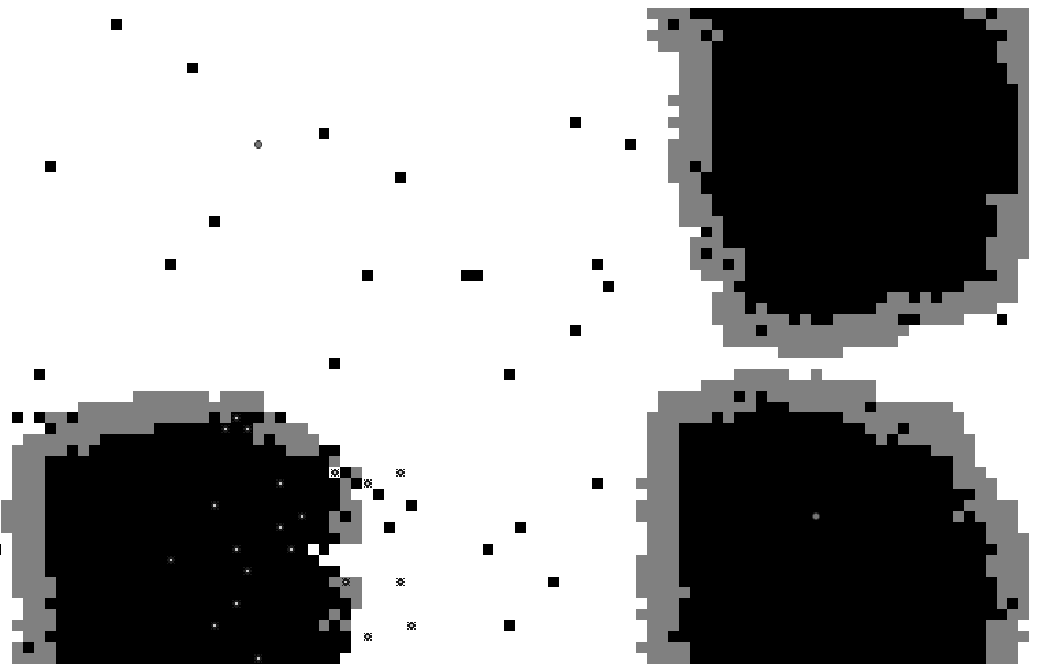}}}
\end{center}
\caption{Partial screenshots of 4 simulations obtained using the same parameter values and initial cell distribution at day 4 post-infection using the global (top row) or local (bottom row) epithelial cell regeneration rule, with the addition of immune cells at random locations (left column) or at the site of recruitment (right column). Healthy epithelial cells are white, infected (containing+expressing+infectious) epithelial cells are grey, dead epithelial cells are black, and immune cells are represented as dark circles with a light grey centre. The top left panel corresponds to the original model presented in \cite{cbeau05}.} 
\label{g_sshot}
\end{figure*}

The proliferation of immune cells in the model presented in \cite{cbeau05} was such that when an activated immune cell moved onto an expressing or infectious cell, a new activated immune cell is added with a probability $r_M = 0.25$ at a random location on the grid. The addition of immune cells at random locations can be justified biologically by the scenario of immune cells being activated and proliferating in the lymph nodes, travelling to the site of infection, and surfacing at random locations throughout the infected tissue. But immune expansion could instead be modelled by adding new activated immune cells on the site where the recruiting activated immune cell is located, hence mimicking immune cell (T cell, macrophages, etc.) division at the infection site. This scenario could correspond to immune cells being activated in the lymph nodes, but travelling to the site of infection while still undergoing their programmed cycles of divisions. In Figure \ref{g_simres}, the infection dynamics for the addition of immune cells at random locations and at the site of recruitment are compared for the two choices of epithelial cell regeneration rule.
A typical spatial distribution of cells at day 4 post-infection under the two immune cell proliferation rules for both epithelial cell regeneration rules are illustrated in Figure \ref{g_sshot} as screenshots of the simulation grid.
Additionally, the numbers of infected and dead epithelial cells at their respective peaks for all rules relative to their values in the original model introduced in \cite{cbeau05} are presented in Table \ref{t_compare}.
\begin{table*}
\begin{tabular}{cccc}
\hline
\hline
\parbox{0.2\linewidth}{epithelial cell regeneration occurs} & \parbox{0.2\linewidth}{newly recruited immune cells placed at} & \parbox{0.2\linewidth}{maximum infected cells (relative)} & \parbox{0.2\linewidth}{maximum dead cells (relative)} \\
\hline
& random locations	& 1.0	& 1.0 \\
\raisebox{1.5ex}[0pt]{globally}
& recruitment site	& 1.1	& 0.46 \\
\hline
& random locations	& 0.80	& 2.1 \\
\raisebox{1.5ex}[0pt]{locally}
& recruitment site	& 0.85	& 2.3 \\
\hline
\hline
\end{tabular}
\caption{The effects of the epithelial cell regeneration rules and the immune cell recruitment rules on the number of infected and dead cells at their respective peak. The numbers are relative to their values for the rules of the original model introduced in \cite{cbeau05}, namely global epithelial cell regeneration with the addition of immune cells at random locations.}
\label{t_compare}
\end{table*}

Regardless of the epithelial cell regeneration rule, the addition of immune cells at the site of recruitment results in more infected cells at the peak of the infection than addition at random locations. The addition of immune cells at random locations allows recruited immune cells to surface randomly onto a previously unexplored site and efficiently discover new patches of infection. With the addition of immune cells at the site of recruitment, it takes longer for immune cells to discover new sites of infection as they can only find them by diffusion. Thus, although the discovered infection sites are cleared faster and more efficiently with the addition of immune cells at the site of recruitment, the undiscovered infection sites are allowed to grow for longer, resulting in more infected cells overall.

In contrast, the addition of immune cells at the site of recruitment rather than at random locations has a different impact on the number of dead cells at the peak for the two epithelial cell regeneration rules. The addition of immune cells at the site of recruitment results in fewer dead cells when combined with the global epithelial cell regeneration rule, but more dead cells when combined with the local regeneration rule. This discrepancy in the effects of the choice of immune cell addition rule for the two epithelial cell regeneration rules can be explained as follows. For the global epithelial cell regeneration rule, the addition of immune cells at random locations allows the infection to grow almost undisturbed while the immune cells slowly populate the grid randomly through recruitment, mainly landing on healthy sites. But when a sufficient number of immune cells have been added, that new immune cells tend to be placed on infected sites, the destruction of infected cells by immune cells begins and happens very abruptly. It is this abrupt destruction of infected cells by immune cells that results in the greater number of dead cells seen with the addition of immune cells at random locations rather than at the site of recognition with the global epithelial cell regeneration rule. This also happens when using the local epithelial cell regeneration rule, but in this case the effect is masked by the large increase in cell destruction at undiscovered infection sites.

In fact, with the addition of immune cells at the site of recruitment and the local epithelial cell regeneration rule, the undiscovered site are sometimes allowed to grow to such extent that the infection gets cleared by target-cell limitation in those areas.

\subsection{Occurrence of chronic infection}

\begin{figure*}
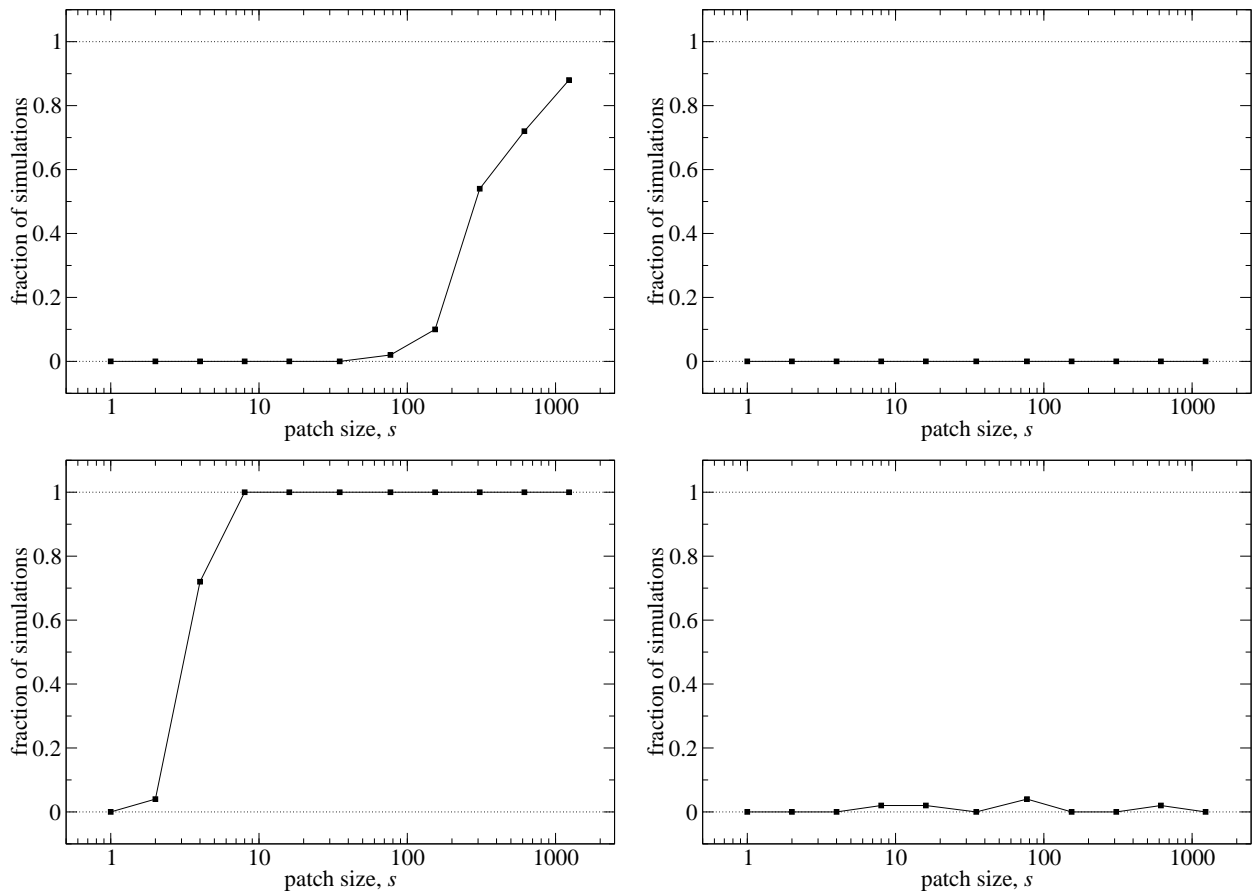

\begin{center}
\resizebox{0.45\linewidth}{!}{\includegraphics{chronic_infec}}
\hspace{0.01\linewidth}
\resizebox{0.45\linewidth}{!}{\includegraphics{chronic_on_mom}} \\
\vspace{0.01\linewidth}
\resizebox{0.45\linewidth}{!}{\includegraphics{chronic_local}}
\hspace{0.01\linewidth}
\resizebox{0.45\linewidth}{!}{\includegraphics{chronic_local_on_mom}}
\end{center}
\caption{Fraction of simulations ending in chronic infection as a function of the initial patch size, using the global (top row) or local (bottom row) epithelial cell regeneration rule, with the addition of immune cells at random locations (left column) or at the site of recruitment (right column). The results were obtained by averaging over 50 simulation runs.}
\label{g_chronic}
\end{figure*}

Examination of the runs in which immune cells are added at the site of recruitment rather than at random locations reveals a dramatic decrease in the fraction of simulations ending in chronic infection. The addition of immune cells at the site of recruitment using the global epithelial cell regeneration rule produced no chronic infection in any of the 50 simulations performed for each initial patch size. Using the local epithelial cell regeneration rule, the addition of immune cells at the site of recruitment produced only a handful of simulations resulting in chronic infection, with the fraction of infected cells stabilizing at approximately $0.1\%$ in all cases. This is illustrated in Figure \ref{g_chronic}. 

The reduction in the fraction of simulations resulting in chronic infection when adding immune cells at the site of recruitment rather than at random locations is easily explained. At high infection levels, the addition of immune cells at the site of recruitment increases the efficacy of the response at the site of recruitment but makes it harder for immune cells to find other sites of infection. This results in a greater number of infected cells. But at low infection levels, immune cells added at random locations will rarely be added at an infection site and are likely to die of old age before they can diffuse to an escaped infection foyer. Thus, the addition of immune cells at random locations is the better strategy for high levels of infection allowing rapid detection of the various infection sites, while addition at the recruitment site is the better strategy for low levels of infection allowing efficient prevention of escape.

\section{In the context of influenza A}
\label{sec_flu}

Influenza is a good example of a spatially localized viral infection. The infection typically takes place in the upper sixteen generations of the lungs, and the target cells of the infection, the ciliated epithelial cells which cover the respiratory tract, are fixed in place. In previous work \cite{cbeau05}, the CA model used here was introduced and successfully calibrated to mimic a viral infection with influenza A. Here, I revisit the CA model to explore how the local epithelial cell regeneration rule and the immune cell addition rule affect the agreement between the CA model and the experimental data cited in \cite{cbeau05} for an uncomplicated influenza A viral infection.

Because the target cells of influenza A are fixed, i.e.\ do not move around in space, it is ultimately the speed of diffusion of the virions over the epithelial layer which determines whether the population of infected cells grows locally around a productively infected cell, or in a more homogeneous manner as the virions quickly spread out over the target area. But since the lifespan of a productively infected cell, the number of virions it produces, and their clearance rate are not well known in the case of influenza A, it is not possible to compute with confidence the rate of spread of virions, and consequently, no conclusion can be made about the true impact of the spatial distribution of cells on the particular development and outcome of an influenza infection. However, there are still some conclusions to be drawn from the results presented above.
 
Originally, in \cite{cbeau05}, the use of a global epithelial cell regeneration rule seemed appropriate to mimic the replacement of dead cells by basal cells or by cells from inferior epithelial layers. I have since learned that the airway epithelium consists of a single layer of cells (everywhere except in the trachea) \cite{virology5th}. Thus, it would seem that a local regeneration rule by which a dead epithelial cell is replaced by a healthy cell only if one of its healthy neighbours divides is more appropriate to model cellular regeneration following a viral infection in the lungs. As it turns out, the use of the local epithelial cell regeneration rule does in fact improve the fit of the CA model to available experimental data. Over the course of an influenza infection, there should be about 10\% of cells dead on day 1, 40\% on day 2 and 10\% on day 5 \cite{bocharov94}. The global rule results in too fast a regeneration, but the local rule improves the agreement of the number of dead epithelial cells during regeneration.

The local epithelial cell regeneration rule also results in a number of infected cells at the peak of the infection ($\sim 40\%$ of the total) which is smaller than that obtained with the global regeneration rule ($\sim 50\%$ of the total). Unfortunately, there is no data available to assess whether the reduction in the number of infected cells at the peak of the infection constitutes an improvement of the model or not. The other two existing mathematical models of influenza A, which are ODE models, have arrived at numbers of infected cells at infection peak of 40\%--78\% \cite{baccam05}, and 60\%--80\% \cite{bocharov94} of the total. Experimental data about the fraction of cells infected at the peak of the infection would therefore be invaluable in discriminating between the different models for influenza A and help determine whether spatial heterogeneity plays a role in the development and outcome of the infection.

Finally, it has been suggested in \cite{baccam05} that influenza resolution could be target-cell limited. This means that the infection would die from the lack of new cells to infect, rather than as a result of immune attacks. In a spatially homogeneous model, the target-cell limited resolution of the infection would be difficult to show \cite{baccam05}. In the context of a spatial model, however, a target-cell limited resolution of the infection can easily be explored as it would simply correspond to a lack of spatially accessible target cells for infectious cells to infect.

With the model in its current state, target-cell limitation can occur locally, as seen using the local epithelial cell regeneration rule with the addition of immune cells at the site of recruitment (see Section \ref{sec_imm}). Resolution of the infection through target-cell limitation alone, however, is not possible. In the absence of immune cells, target-cell limitation is such that sites of infection grow undisturbed and as the circular waves of infection meet and annihilate, they leave behind nothing but dead cells. Target-cell limited resolution is impossible because as long as the infection wave encircles the dead epithelial cells, segregating them from healthy cells, regeneration cannot be initiated. It is only once immune cells have started attacking the propagating infection wave, creating breaks where dead cells can be in contact with healthy cells, that epithelial cell regeneration can take place. In the absence of immune cells, target-cell limited resolution could still be explored with our model, provided that the action of interferon be included. Type I interferon (IFN) get produced during an influenza infection and hinder viral replication within infected cells and confer a certain level of protection from infection in surrounding cells \cite{baccam05}. In the CA model, the interferon response could be modelled by introducing an inhomogeneous infection rate or an infection rate that would depend on the number of infectious neighbours. This could be the subject of future research.

\section{Conclusion}

Here, the CA model introduced in \cite{cbeau05} was used to investigate the effects of the well-mixed assumption on the dynamics of a localized viral infection. It was shown that the distribution of initially infected cells has a great impact on the dynamics of infection. This is because, in the CA model, infectious cells can only infect their immediate neighbours, and when organized in patches, fewer infectious cells have healthy neighbours.

It was also demonstrated that the regeneration rule chosen for the replacement of dead epithelial cells by healthy ones can have an important impact on infection dynamics. A global epithelial cell regeneration rule, as is equivalent to simple ODE models, allows areas of dead cells to be replenished by healthy cells even in the local absence of healthy cells. This repopulation, in turn, allows the infection to move back into the newly replenished area it had previously infected, resulting in a greater number of infected cells. On the other hand, the slower local regeneration rule, which requires the local presence of healthy epithelial cells, limits the growth of the infection by starving it of target cells and forces the infection to propagate as a thin circular wave. In \cite{strain02}, Strain et al.\ introduce a spatiotemporal model for the dynamics of HIV in the spleen. Strain et al.\  point out that the main differences between their spatial model and a mean field approach such as an ODE model, arise from the fact that a viral burst only spreads to nearby cells. They also conclude that in a spatial model, infection sustainability is affected by the recovery rate of destroyed target cells, as local cell destruction limits the spread of the infection which can then only be sustained as a propagating wave. Those findings are in agreement with those presented here.

Finally, the choices of whether to add immune cells at random locations on the simulation grid, as is equivalent to simple ODE models, or at the site of recruitment were compared to explore how they affect the dynamics of the infection. It was shown that while addition at random sites permits rapid detection of new infection sites, it makes it harder to avoid infection escape from the immune response. Consequently, random addition of immune cells was found to be a better strategy at high infection levels, while addition at the site of recruitment was the better strategy at low infection levels.

Two spatial models \cite{zorzenon01,strain02} have been suggested for the dynamics of HIV infections. Both models make the assumption that T cells, the target cells of HIV virus, are fixed in space, an assumption that is not realistic given the known patterns of movement of T cells within lymph nodes \cite{miller02,miller03,miller04jem,miller04pnas,mempel04}, and may adversely affect the results. Other investigations \cite{funk05,louzoun01} have chosen to remain more general in their exploration of the effects of the spatial distribution of agents on the evolution and outcome of infections by not considering a particular viral infection. Since the models in \cite{funk05,louzoun01} have not been calibrated to fit experimental data, it is not known whether they can realistically model any particular infection. In \cite{cbeau05}, the model used here was calibrated for influenza A, and was shown to be accurate enough to quantitatively reproduce the response to an uncomplicated infection with this virus. The applicability of the findings presented here follow from that model.

In the present work, the effect of the spatial distribution of infected cells on the dynamics of the infection arises from the fact that the infection can only spread from one infectious cell to its neighbours. The applicability of the findings presented here largely depends on the accuracy of this assumption, namely whether the infection tends to quickly spread over the tissue or grow locally around infected sites. Nonetheless, I have shown in this paper that a local epithelial cell regeneration rule, where a dead cell is replaced by a healthy cell when one of its immediate healthy neighbour divides, improves the fit of the CA model to experimental data in the case of an uncomplicated viral infection with influenza A.

Whether or not there exist \latin{in vivo} virus-host systems where the infection grows locally from neighbour to neighbour, such systems do exist \latin{in vitro} and are used to address questions such as how viral spread is inhibited by cellular antiviral activities \cite{duca01,lam05}. The team of Dr.\ John Yin, at the University of Wisconsin-Madison, have introduced a new assay method which consists of a monolayer cell culture covered in an agar solution, which prevents the diffusion of virions at the surface of the cell monolayer such that the infection can only spread to immediate neighbours \cite{duca01,lam05}, as is the case in the CA model used here. By complementing these assay experiments with simulations from the CA model used here, significant questions could be addressed. For example, by testing various hypothesis about the production and spread of interferon, and comparing the results of the CA model to that of the experimental assays, it may be possible to discriminate among various potential mechanisms and extract parameters for those mechanisms, such as rate of production/clearance of interferon. The combination of results obtained through such experimental techniques with the flexibility and simplicity offered by spatial \latin{in silico} modelling could lead to great advances in our understanding of host-pathogen interactions.

\begin{acknowledgments}
This work was supported in part by MITACS' Mathematical Modelling in Pharmaceutical Development (MMPD) project. Portions of this work were done under the auspices of the U.S.\ Department of Energy under contract W-7405-ENG-36 and supported by NIH grant AI 28433 awarded to Dr.\ Alan S.\ Perelson. The author also wishes to thank Dr.\ Kipp Cannon and Dr.\ Alan S.\ Perelson for helpful discussions.
\end{acknowledgments}

\appendix*

\section{Notation}

Table \ref{t_notation} lists the relationship between the notation used in this document and that of \cite{cbeau05}.

\begin{table*}
\begin{center}
\begin{tabular}{ccl}
\hline
\hline
Notation in \cite{cbeau05} & New notation & Description \\
\hline
\texttt{FLOW\_RATE} & $\nu$ & Speed of immune cells \\
\texttt{IMM\_LIFESPAN} & $\delta_M$ & Lifespan of an immune cell \\
\texttt{CELL\_LIFESPAN} & $\delta_H$ & Lifespan of healthy epithelial cells \\
\texttt{INFECT\_LIFESPAN} & $\delta_I$ & Lifespan of infected epithelial cells \\
\texttt{INFECT\_INIT} & $\rho_C$ & Proportion of initially infected cells \\
\texttt{INFECT\_RATE} & $\beta$ & Rate of infection of neighbours \\
\texttt{EXPRESS\_DELAY} & $\tau_E$ & Delay from containing to expressing \\
\texttt{INFECT\_DELAY} & $\tau_I$ & Delay from containing to infectious \\
\texttt{DIVISION\_TIME} & $b$ & Duration of epithelial cells' division \\
\texttt{BASE\_IMM\_CELL} & $\rho_M$ & Minimum density of immune cells per epithelial cell \\
\texttt{RECRUITMENT} & $r_M$ & Number of immune cells recruited when one recognizes the virus \\
New parameter & $s$ & Size of patches of initially infected cells \\
\hline
\hline
\end{tabular}
\end{center}
\caption{Relationship between the notation used in this document and that of \cite{cbeau05}.}
\label{t_notation}
\end{table*}

\addcontentsline{toc}{section}{References}
\bibliographystyle{abbrv}
\bibliography{../../../../local/TeX/allbibliographies}

\begin{thebibliography}{10}

\bibitem{baccam05}
P.~Baccam, C.~Beauchemin, C.~A. Macken, F.~G. Hayden, and A.~S. Perelson.
\newblock Kinetics of influenza {A} virus infection in humans.
\newblock {\em In Preparation}, 2005.

\bibitem{cbeau05}
C.~Beauchemin, J.~Samuel, and J.~Tuszynski.
\newblock A simple cellular automaton model for influenza {A} viral infections.
\newblock {\em J. Theor. Biol.}, 232(2):223--234, 21 January 2005.
\newblock Draft available on arXiv:q-bio.CB/0402012.

\bibitem{bocharov94}
G.~A. Bocharov and A.~A. Romanyukha.
\newblock Mathematical model of antiviral immune response {III}. {I}nfluenza
  {A} virus infection.
\newblock {\em J. Theor. Biol.}, 167(4):323--360, 1994.

\bibitem{duca01}
K.~A. Duca, V.~Lam, I.~Keren, E.~E. Endler, G.~J. Letchworth, I.~S. Novella,
  and J.~Yin.
\newblock Quantifying viral propagation \emph{in vitro}: Towards a method for
  characterization of complex phenotypes.
\newblock {\em Biotechnol. Prog.}, 17(6):1156--1165, December 2001.

\bibitem{durrett94}
R.~Durrett.
\newblock The importance of being discrete (and spatial).
\newblock {\em Theor. Popul. Biol.}, 46(3):363--394, December 1994.

\bibitem{durrett-levin94}
R.~Durrett and S.~A. Levin.
\newblock Stochastic spatial models: A user's guide to ecological applications.
\newblock {\em Philosophical Transactions: Biological Sciences},
  343(1305):329--350, 1994.

\bibitem{funk05}
G.~A. Funk, V.~A. Jansen, S.~Bonhoeffer, and T.~Killingback.
\newblock Spatial models of virus-immune dynamics.
\newblock {\em J. Theor. Biol.}, 233(2):221--236, 21 March 2005.

\bibitem{hagenaars04}
T.~J. Hagenaars, C.~A. Donnelly, and N.~M. Ferguson.
\newblock Spatial heterogeneity and the persistence of infectious diseases.
\newblock {\em J. Theor. Biol.}, 229(3):349--359, 2004.

\bibitem{lam05}
V.~Lam, K.~A. Duca, and J.~Yin.
\newblock Arrested spread of vesicular stomatitis virus infections in vitro
  depends on interferon-mediated antiviral activity.
\newblock {\em Biotechnol. Bioeng.}, 90(7):793--804, 30 June 2005.

\bibitem{lloyd96}
A.~L. Lloyd and R.~M. May.
\newblock Spatial heterogeneity in epidemic models.
\newblock {\em J. Theor. Biol.}, 179(1):1--11, 1996.

\bibitem{louzoun01}
Y.~Louzoun, S.~Solomon, H.~Atlan, and I.~R. Cohen.
\newblock Modeling complexity in biology.
\newblock {\em Physica A}, 297:242--252, 2001.

\bibitem{mempel04}
T.~R. Mempel, S.~E. Henrickson, and U.~H. {von Adrian}.
\newblock T-cell priming by dendritic cells in lymph nodes occurs in three
  distinct phases.
\newblock {\em Nature}, 427(6970):154--159, 8 January 2004.

\bibitem{miller04pnas}
M.~J. Miller, A.~S. Hejazi, S.~H. Wei, M.~D. Cahalan, and I.~Parker.
\newblock {T} cell repertoire scanning is promoted by dynamic dendritic cell
  behavior and random {T} cell motility in the lymph node.
\newblock {\em P. Natl Acad. Sci. USA}, 101(4):998--1003, 27 january 2004.

\bibitem{miller04jem}
M.~J. Miller, O.~Safrina, I.~Parker, and M.~D. Cahalan.
\newblock Imaging the single cell dynamics of {CD4}$^+$ {T} cell activation by
  dendritic cells in lymph nodes.
\newblock {\em J. Exp. Med.}, 200(7):847--856, 2004.

\bibitem{miller03}
M.~J. Miller, S.~H. Wei, M.~D. Cahalan, and I.~Parker.
\newblock Autonomous {T} cell trafficking examined \emph{in vivo} with
  intravital two-photon microscopy.
\newblock {\em P. Natl Acad. Sci. USA}, 100(5):2604--2609, 2003.

\bibitem{miller02}
M.~J. Miller, S.~H. Wei, I.~Parker, and M.~D. Cahalan.
\newblock Two-photon imaging of lymphocyte motility and antigen response in
  intact lymph node.
\newblock {\em Science}, 296(5574):1869--1873, 7 June 2002.

\bibitem{perelson02}
A.~S. Perelson.
\newblock Modelling viral and immune system dynamics.
\newblock {\em Nature Rev. Immunol.}, 2(1):28--36, 2002.

\bibitem{perelson96hiv}
A.~S. Perelson, A.~Neumann, M.~Markowitz, J.~Leonard, and D.~Ho.
\newblock {HIV}-1 dynamics \emph{in vivo}: virion clearance rate, infected cell
  life-span, and viral generation time.
\newblock {\em Science}, 271:1582--1586, 1996.

\bibitem{virology5th}
C.~W. Potter.
\newblock Influenza.
\newblock In A.~J. Zuckerman, J.~E. Banatvala, J.~R. Pattison, P.~D. Griffiths,
  and B.~D. Schoub, editors, {\em Principles and Practice of Clinical
  Virology}, chapter~5, pages 271--297. John Wiley \& Sons, Ltd., 5th edition,
  2004.

\bibitem{strain02}
M.~C. Strain, D.~D. Richman, J.~K. Wong, and H.~Levine.
\newblock Spatiotemporal dynamics of {HIV} propagation.
\newblock {\em J. Theor. Biol.}, 218(1):85--96, 7 September 2002.

\bibitem{young01}
W.~R. Young, A.~J. Roberts, and G.~Stuhne.
\newblock Reproductive pair correlations and the clustering of organisms.
\newblock {\em Nature}, 412:328--331, 19 July 2001.

\bibitem{zorzenon01}
R.~M. {Zorzenon dos Santos} and S.~Coutinho.
\newblock Dynamics of {HIV} infection: A cellular automata approach.
\newblock {\em Phys. Rev. Lett.}, 87(16), 15 October 2001.

\end{thebibliography}

\end{document}